# Quantum interference engineering of nanoporous graphene for carbon nanocircuitry

Gaetano Calogero,[‡ab†] Isaac Alcón,[*‡ab§] Nick Papior,[bc] Antti-Pekka Jauho[ab] and Mads Brandbyge[*ab]

[a] Department of Physics, Technical University of Denmark, DK-2800 Kongens Lyngby, Denmark

[b] Center for Nanostructured Graphene (CNG), DK-2800 Kongens Lyngby, Denmark

[c] Computing Center, Technical University of Denmark, DK-2800 Kongens Lyngby, Denmark

KEYWORDS: nanoporous graphene, quantum interference, nanocircuitry, bond currents, equilibrium Green's function, tight-binding, density-functional theory, materials design

**ABSTRACT:** Bottom-up prepared carbon nanostructures appear as promising platforms for future carbon-based nano-electronics, due to their atomically precise and versatile structure. An important breakthrough is the recent preparation of nanoporous graphene (NPG) as an ordered covalent array of graphene nanoribbons (GNRs). Within NPG, the GNRs may be thought of as 1D electronic nanochannels through which electrons preferentially move, highlighting NPG's potential for carbon nanocircuitry. However, the π-conjugated bonds bridging the GNRs give rise to electronic cross-talk between the individual 1D channels, leading to spatially dispersing electronic currents. Here, we propose a chemical design of the bridges resulting in destructive quantum interference, which blocks the cross-talk between GNRs in NPG, electronically isolating them. Our multiscale calculations reveal that injected currents can remain confined within a single, 0.7 nm wide, GNR channel for distances as long as 100 nm. The concepts developed in this work thus provide an important ingredient for the quantum design of future carbon nanocircuitry.

## Introduction

Bottom-up on-surface synthesis of carbon-based nanostructures has been undergoing an important development for more than a decade.[1–3] In this approach, specifically designed molecular building blocks are deposited on metallic substrates where they self-assemble and react generating atomically precise, ordered and robust (covalent) 2D nano-structures.[3] Due to the high versatility of organic synthesis very diverse nanostructures have been reported, such as graphene nanoribbons[4] (GNRs), nano-graphenes[5] and 2D covalent organic frameworks (2D-COFs) of multiple topologies.[2] Their atomically precise character allows accommodating subtle nano-electronic phenomena, such as the recently reported topological quantum states in GNRs,[6,7] which are otherwise very difficult to obtain via alternative top-down nanostructuring approaches. Despite the fact that these ground breaking advances are expected to play a central role in future carbon-based nano-electronics,[2] to date it is still not clear how to transform all this structural and electronic versatility into specific technological functions.

The recent bottom-up synthesis of nanoporous graphene (NPG) composed of laterally bonded atomically precise graphene nanoribbons[8] (GNRs) represents an important breakthrough in this direction. As shown in **Fig. 1a**, the nanoribbons are covalently integrated in a highly ordered and robust monolayer, which allowed transferring the entire two-dimensional (2D) array onto an insulating substrate necessary for solid-state devices.[8] Thus GNRs within NPG may be thought of as 1D channels through which electrons preferentially move, which makes NPG's structure particularly appealing for nanocircuitry: i.e. for directing electrons through pre-designed paths, which is a core component of any electronics technology. However, large-scale atomistic simulations have recently shown that currents injected into single GNRs in NPG are rapidly spatially dispersed due to the electronic coupling between the 1D channels.[9] This is caused by the conducting π-conjugated carbon-carbon bonds connecting GNRs (see inset and model in **Fig. 1a**) and severely challenges the usage of NPGs of this type as a platform for nanocircuitry.

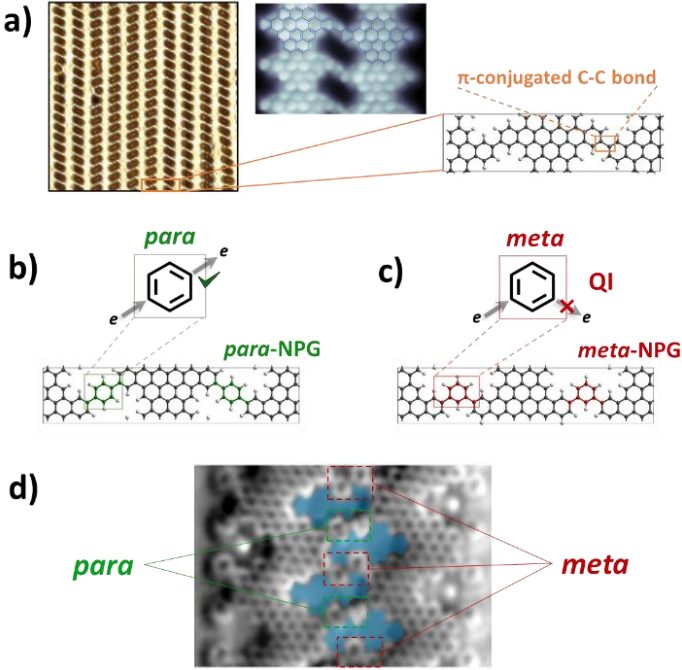

**Fig. 1. Proposal of *para* and *meta* bridges in NPG. a)** Scanning tunnelling microscopy (STM) image of the bottom-up prepared NPG (20 nm x 20 nm) composed of individual GNRs laterally connected (see high-resolution STM image in the inset and atomistic model). From Ref.4; reprinted with permission from AAAS. **b)** *Para* and **c)** *meta* connections through a benzene ring leading to a high transmission and suppressed transmission due to destructive quantum interference (QI), respectively. Bottom panels in **b)** and **c)** show the atomistic models' unit cells of our proposed NPGs, highlighting *para* benzene bridges in green and *meta* benzene bridges in red, respectively. Remaining atoms are coloured as C – grey and H – white. **d)** STM image of recently synthesized GNRs connected with a combination of *para* and *meta* benzene bridges. Adapted with permission from Ref. 15. Copyright 2018 American Chemical Society.

Destructive quantum interference (QI) is known to cause a dramatic decrease in electronic transmittance through nanoscale devices such as single-molecule break junctions.[10,11] QI has been correlated with fundamental principles of π-conjugation.[12] For instance, considering benzene as a simple nanowire, electronic transmission takes place if electrodes are contacted in *para* positions (**Fig. 1b**). On the contrary, if electrodes are contacted in *meta* positions (**Fig. 1c**) QI occurs leading to a significant suppression of electrical conductance.[10,11,13,14] Due to the π-conjugated nature of NPG,[8] bonding GNRs within it via *para* or *meta* connections through benzene bridges is a promising strategy to tune the electronic coupling between them. This should in turn allow for engineering electrical current paths within the 2D material.

In this work, we present a novel QI-based design of NPG, by bonding GNRs within it via benzene bridges with either *para* or *meta* connections, resulting in the *para*-NPG and *meta*-NPG structures, respectively (bottom panels in **Fig. 1b-c**). Our results, based on multi-scale atomistic simulations, show that the electronic coupling between individual GNR 1D channels depends on the particular type of connection (*para* or *meta*) due to QI. This allows engineering electronic currents injected in these NPGs, which may spatially disperse over a number of GNRs as they propagate (*para*-NPG), such as reported[9] for the recently fabricated NPG,[8] or may be confined within a single GNR channel for distances longer than 100 nm (*meta*-NPG). Recently, bottom-up synthesized GNRs have been bonded via benzene-bridges with both *para* and *meta* connections[15] (see **Fig. 1d**), supporting the experimental feasibility of our proposed structures. Below we further explore the electronic tunability offered by *para* and *meta* connections by designing a hybrid *para-meta-para*-NPG where electrons follow more complex paths. Overall, in this work we propose the use of QI-based engineering as a tool to design, and realize, bottom-up graphene nano-structures for future carbon-based nanocircuitry.

## Results

The fabricated NPG (**Fig. 1a**) behaves as an array of weakly coupled 1D electronic channels (i.e. GNRs). Consequently, electrons injected in one of the GNRs spread forming a Talbot interference pattern.[9] In this context, the electronic wave amplitude ($\psi_n$) inside the $n$th GNR, aligned along $y$ within NPG, is the solution to the coupled-mode equation:[16,17]

$$i\frac{d\psi_n}{dy}(y) + \kappa_c\left[\psi_{n-1}(y) + \psi_{n+1}(y)\right] = 0, \quad \text{(Eq. 1)}$$

where $n$ is the index of the particular GNR channel, $y$ the longitudinal position within that channel, and $\kappa_c$ the inter-channel coupling coefficient, determining the degree of cross-talk between GNRs and the subsequent electronic spreading. The $\kappa_c$ parameter can be estimated from the NPG band structure at a particular energy from the momentum difference $\Delta k$ between the first two conduction (or valence) bands, as $\kappa_c = \Delta k/4$.[9] Due to QI, connecting GNRs with benzene bridges in *para* or *meta* appears as an ideal strategy to engineer the inter-channel coupling such that $\kappa_c^{para} >> \kappa_c^{meta}$.

To assess this hypothesis we have designed the two novel *para*-NPG and *meta*-NPG shown in **Fig. 1b-c**. The extra benzene ring connecting GNRs is the main difference with the recently fabricated NPG[8] (**Fig. 1a**). Both *para* and *meta*-NPG structures are modelled using periodic boundary conditions, and their atomic and electronic structure has been optimised with the density functional theory (DFT) approach using the SIESTA package (see Methods).

The *para* and *meta*-NPG have been designed from perfectly planar structures, neglecting out-of-plane distortions during optimization. This has been done in order to isolate the effect of QI from other conformational parameters (e.g. twist angles). However, if small out-of-plane distortions are introduced before optimization, both structures fall into energetic minima where the benzene bridges are slightly out-of-plane (see Supporting Fig. S1). In the Supporting Information we show that the out-of-plane relaxations have minor impact on the electronic structure around the Fermi level, the transport, and our conclusions. Thus, for simplicity, in the following we present results for the planar *para* and *meta*-NPG structures, and point out the role of distortions where relevant.

In order to access the transport properties of the NPG at the 100 nm scale with DFT accuracy (below) we use the open-source in-house developed tools SISL and TBtrans.[18,19] SISL allows us to extract an effective, non-orthogonal, tight-binding (TB) description of the carbon π-system from DFT. This TB parametrized model (Hamiltonian and overlap) of the system, composed of some hundred thousand atoms, is subsequently used to perform transport calculations with TBtrans based on the Green's function (GF)[20,21] formalism (see Methods). We compute all results shown below using this TB model, highlighting differences with the full DFT in the Supporting Information.

**Fig. 2c**-*left* and **2d**-*left* show the TB band structure for *para*-NPG and *meta*-NPG, respectively. These are in good qualitative agreement with those obtained from DFT, except for an energy rescaling (see Supporting Fig. S2). Both *para*-NPG and *meta*-NPG band structures exhibit a band-gap of ≈ 0.7 eV. Furthermore, in both cases the first two conduction (and valence) bands have a strong dispersion along Γ → Y, in agreement with results obtained for the fabricated NPG,[8,9] compared to a small (*para*-NPG) or even negligible (*meta*-NPG) dispersion along Γ → X. Such differences of dispersion along Γ → X give rise to different splitting, $\Delta k$, between the two conduction (or valence) low-energy bands along Γ → Y, being $\Delta k$ significantly larger for the *para*-NPG than for the *meta*-NPG (see red labels in **Fig. 2c-d** left panels). The values of $\Delta k$ present low variations over the low-energy range for both systems, but $\Delta k_{para}$ always remains around one order of magnitude larger than $\Delta k_{meta}$ (see Supporting Fig. S3). The difference in $\Delta k$ indicates that low-energy waves in the *para*-NPG are significantly more coupled compared to the *meta*-NPG. Indeed, by plotting $|\Psi_1 + \Psi_2|$ at E - $E_F$ = 0.7 eV in real-space for 40 nm x 120 nm NPG samples, where $\Psi_1$ and $\Psi_2$ are the two conduction low-energy Bloch states, we observe a modulation along the *y* axis in the *para*-NPG (**Fig. 2c**-*right*), similar to that reported for the fabricated NPG,[9] which is absent in the *meta*-NPG (**Fig. 2d**-*right*). Upon out-of-plane distortions of benzene bridges the qualitative difference between *para* and *meta*-

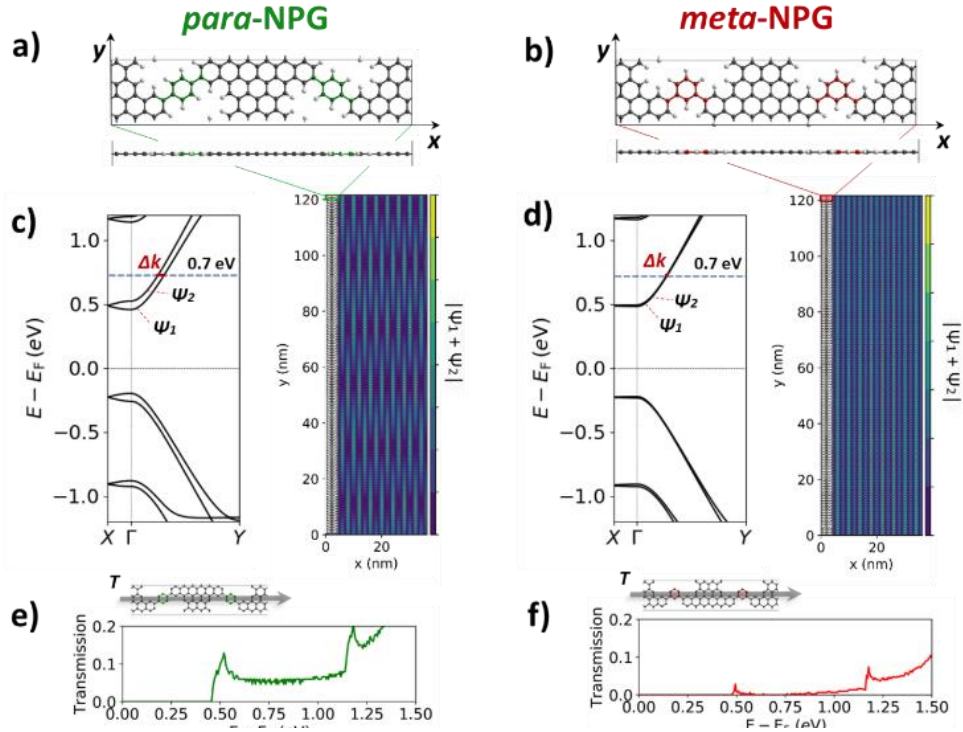

**Fig. 2. Electronic characterization of *para*-NPG and *meta*-NPG. (a-b)** DFT optimized unit cell for *para*-NPG and *meta*-NPG, respectively (top *z*-view; bottom *y*-view), highlighting *para* benzene bridges in green and *meta* benzene bridges in red. Remaining atoms are coloured as C – grey and H – white. **(c-d)** *Left* Band structure of *para*-NPG and *meta*-NPG, respectively, as calculated with a DFT-parameterized TB model. Low-energy conduction Bloch states are indicated as $\Psi_1$ and $\Psi_2$ and their momentum difference at E - $E_F$ = 0.7 eV as $\Delta k$ in red. *Right* Real-space representation of $|\Psi_1 + \Psi_2|$ at E - $E_F$ = 0.7 eV (blue dashed line in band structures) for *para*-NPG **(c)** and *meta*-NPG **(d)** 40 nm x 120 nm samples. **(e-f)** Zero-bias transmission along *x* direction (transverse to GNRs, as sketched on top of each graph) for *para*-NPG and *meta*-NPG, respectively.

NPG band structures is preserved (see Supporting Fig. S4). We note, though, that due to the twist of benzene bridges in the *para*-NPG (Supporting Fig. S1) band splitting, and thus $\kappa_c$, is slightly reduced (see Supporting Fig. S4a-b).

We have investigated transport both along GNRs ($y$-direction) and perpendicular to them ($x$-direction). As expected, electrons mainly propagate along GNR channels rather than through their bridges, with transmission in the $y$ direction being significantly larger than along the $x$ direction over the entire energy spectrum, regardless of the type of NPG (see Supporting Fig. S5). However, focusing on transport along the $x$-direction, we find that electrons' transmission through *para*-bridges (Fig. 2e) is an order of magnitude higher than through *meta*-bridges (Fig. 2f) for all energies from 0.5 to 1.1 eV (low-energy regime). This result, which is robust to out-of-plane distortions of benzene bridges (see Supporting Fig. S6), is in agreement with our band structure results (Fig. 2a-d) and with previous transport measurements in *para* and *meta* coupled benzene single-molecule break junctions,[11] where similar differences in electronic conductance were found.

All results shown in Fig. 2 strongly suggest that, as opposed to the *para*-NPG and the fabricated NPG,[8,9] within the *meta*-NPG the GNRs behave as independent 1D electronic channels due to the QI taking place in *meta* benzene bridges. This, in turn, implies that currents injected in the *meta*-NPG should remain confined within single GNRs: i.e., within channels as narrow as 0.7 nm. To confirm this expectation we construct devices of 80 nm x 120 nm NPG samples connected to electrodes composed of the same NPG at the two terminations along $y$, and using absorbing walls on either side along the $x$ direction to avoid inter-cell electronic communication (see Methods). We model a local injection of currents through a single atom (see Methods) which, experimentally, could be realized using a scanning tunnelling microscope (STM) tip.[22] To study the spatial paths followed by the electronic currents we plot the associated bond currents in the device. Fig. 3 shows the path of electrons injected at y = 0 nm in each device (site marked by a red dot) at different energies. We observe a striking difference between the two types of NPG. In the *para*-NPG (Fig. 3a) currents injected at all energies spread over several GNR channels as they propagate towards the electrode at y = 120 nm, giving rise to the Talbot interference pattern.[9] On the contrary, currents injected in the *meta*-NPG (Fig. 3b) are confined within the individual GNR where they have been injected (i.e. within a 0.7 nm wide channel) for over 100 nm. This is true for both electrons injected in the conduction band (e.g., $E - E_F$ = 0.7 eV) and holes injected in the valence band (e.g., $E - E_F$ = -0.4 eV), confirming the ambipolar nature of the effect. Around $E - E_F$ = 0.5 eV (Fig. 3b) and $E - E_F$ = 1 eV (Supporting Fig. S7) we observe some delocalization to the nearest neighbouring GNRs, due to a small increase of $\Delta k$ (and thus $\kappa_c$) at those energies (see Sup-porting Fig. S3). However, we see that even for those energies currents are still confined within a maximum of 3-5 adjacent GNRs after more than 100 nm from the source (Fig. 3b), as opposed to the *para*-NPG where currents disperse through more than 20 adjacent channels at all tested energies (Fig. 3a). We note that the highly dispersive character of currents within the *para*-NPG at E - $E_F$ = 0.5 eV is a consequence of the large $\Delta k$ at the energetic onset of the conduction bands (see $\Delta k$ plots in the Supporting Fig. S3). Out-of-plane distortions in the *meta*-NPG only cause a minor degradation of confinement for currents injected in the valence band (see Supporting Fig. S8), obtaining strong confinement up to 50 nm. Contrary, currents injected in the conduction band are fully confined for the tested distances (approximately 85 nm).

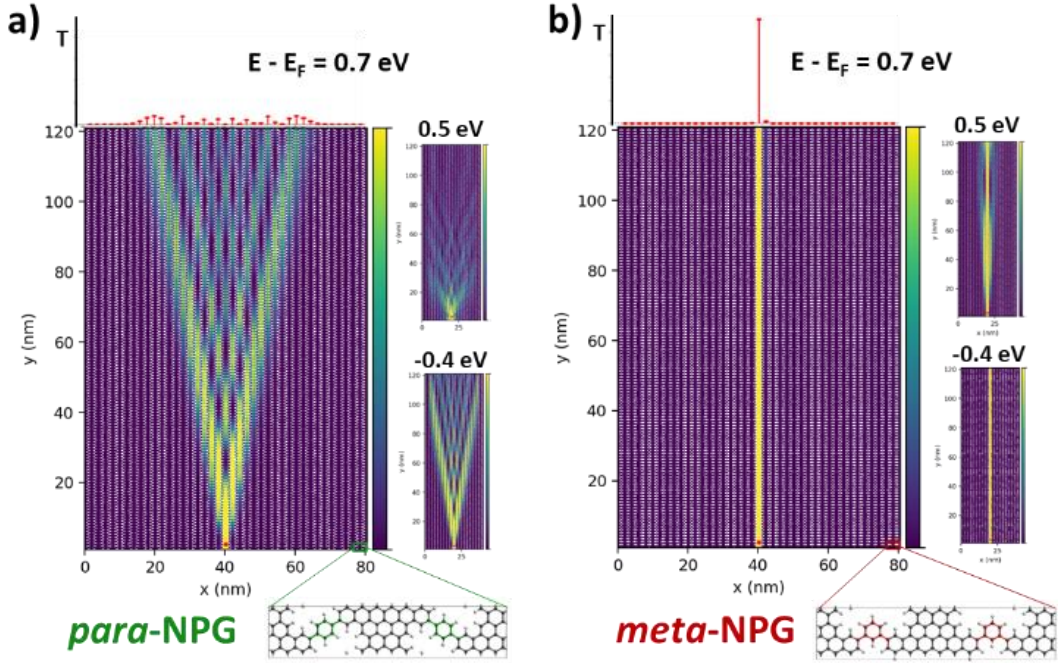

**Fig. 3. Current injection in *para*-NPG and *meta*-NPG.** Bond-currents at $E - E_F$ = 0.7 (large panel), 0.5 and -0.4 eV (small panels) from a large-scale DFT-parameterized TB model (257,600 atoms) of (**a**) *para*-NPG and (**b**) *meta*-NPG, injected on a single site with $(x, y) \approx$ (40 nm, 2 nm) as indicated with the red dot. Each colour map is normalized to its maximum value of current. On top of 0.7 eV graphs the sum of all bond-currents (i.e. total transmission - T) per GNR channel at y = 120 nm is plotted.

These currents should be detectable via dual-probe measurements,[9] where a second STM tip may be used to measure currents while scanning the sample along the *x* direction.[23] On top of the maps at E - $E_F$ = 0.7 eV in **Fig. 3**, we plot the total transmission per GNR channel close to the electrode at y = 120 nm, calculated as the sum of bond currents at that position from the point of injection. Within the *para*-NPG (**Fig. 3a**), due to spreading of currents, the electronic signal per GNR is rather low. On the contrary, within the *meta*-NPG currents are concentrated in a single GNR (**Fig. 3b**) yielding an electronic signal around ten times larger than for the most intense channels in the *para*-NPG. This result strongly supports the experimental feasibility of tracking currents injected in the *meta*-NPG.

The recent developments in the bottom-up fabrication of graphene nano-structures support the experimental feasibility of fabricating *para* and *meta*-NPG by using *para* and *meta* benzene bridges to covalently bond GNRs (see **Fig. 1d**).[15] Other studies have shown that it is possible to fabricate GNRs with alternating widths[6,7] or even covalently merge different types of GNRs using specifically designed interface building blocks.[24] Since NPG consists of laterally bonded GNRs, all these developments in GNR engineering suggest that NPGs with more complex and heterogeneous structures may be realised in the near future, offering novel nano-electronic functionalities. In this respect, the *para*/*meta* control demonstrated here for NPG may be exploited as a tool to design complex electronic nano-paths within a single hybrid NPG monolayer. Since the *para* and *meta*-NPG share the same unit cell parameters one may envisage a hybrid NPG where *para* and *meta*-NPG areas, or "modules", are covalently merged. To explore this idea we designed a hybrid system where a *para*-NPG "module" is inserted between two *meta*-NPG "modules". Electron transport in this more involved system was investigated using the standard orthogonal nearest-neighbour TB parameterization for graphene (on-site energy $\varepsilon_i$ = 0 eV, hopping $t_{ij}$ = -2.7 eV, see Supporting Fig. S9). The DFT-optimized atomic structures for the interfaces used to merge the different modules ($i_{m \to p}$ and $i_{p \to m}$) are available in the Supporting Fig. S10. In **Fig. 4** we show the propagation of injected bond currents through the *meta-para-meta* modules in this hybrid NPG.

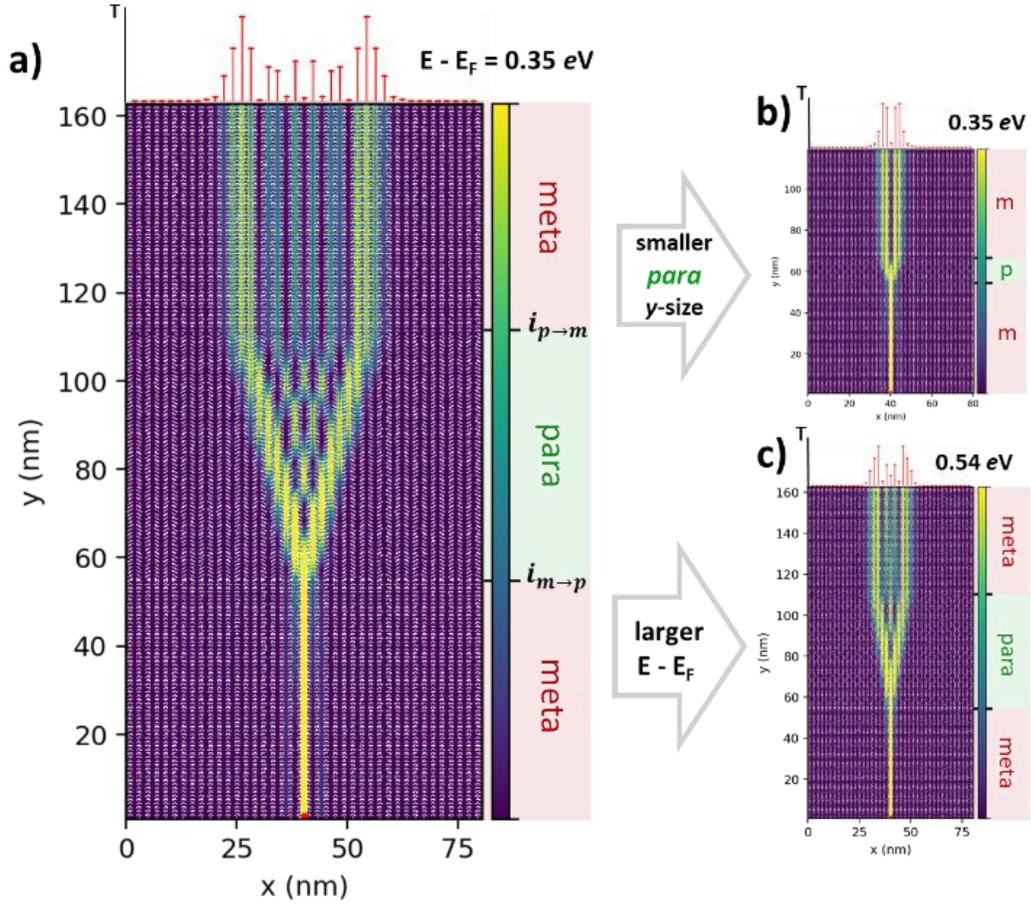

**Fig. 4. Current injection in a hybrid *meta-para-meta*-NPG.** (a) Bond-currents at E - $E_F$ = 0.35 eV in a large-scale (345,920 atoms) hybrid NPG system composed of covalently connected *meta*, *para* and *meta*-NPG modules (each of size 75 × 55 nm). Current is injected at the red dot in the bottom of the frame. (b) Bond-currents at E - $E_F$ = 0.35 eV for a device where the *para*-NPG module is six times shorter along *y* as compared to (a). (c) Bond-currents at E - $E_F$ = 0.54 eV within the same device as in (a). Each colour map is normalized to its maximum value of current. Total transmission (T) per GNR channel recorded at *y* = 160 nm is plotted on top of each device.

As shown in **Fig. 4**, currents injected in the *meta-para-meta*-NPG present a complex pathway, but following the logic found above for the pristine materials. Focusing on **Fig. 4a**, we see that currents injected at the lower *meta*-module propagate confined in a single GNR, in agreement with results of **Fig. 3b**. Then, after approximately 55 nm the localized currents hit interface $i_{m \to p}$ and enter the middle *para*-module, where they spread forming the Talbot interference pattern, in accordance with the results of **Fig. 3a**. We note that at interface $i_{m \to p}$ minor electronic reflection takes place, similar to that reported for GNR-based beam splitters.[25] Next, the currents within the *para*-module reach interface $i_{p \to m}$ and enter the upper *meta*-module. There, due to the electronic decoupling of GNRs via QI, the dispersed currents get channelled again into individual GNRs, which guide the carriers up to the electrode at *y* = 160 nm without any further transverse spreading, thus "freezing" the Talbot pattern generated in the *para*-module. The "frozen" electronic pattern depends on the vertical size of the middle *para*-section. For example, as shown in **Fig. 4b**, making the *para*-module six times shorter gives rise to a beam splitting effect.[26] Additionally, by exploiting the energy dependence of the Talbot spreading angle,[9] increasing the energy of injected electrons allows confining them in a smaller number of GNR channels in the higher *meta*-module (compare **Fig. 4a** and **4c**). This could be done, for instance, using standard electrostatic gating fixing $E_F$ at a specific value. Hence, this type of *para-meta* hybrid NPGs may serve in the future as unique platforms to externally manipulate, via gating, electronic paths with nanometric precision. We note, finally, that as for the pristine NPGs (**Fig. 3**) the tunability of the electronic currents may be read by measuring the total transmission per GNR channel at the end of the device (see top red plots in **Fig. 4a-c**) using, for instance, an STM tip.[9,23]

## Conclusions

We propose a QI-based design of NPG which allows quantum engineering electrical currents at the nanoscale.

We design two new NPGs where GNRs are laterally connected either in *para* or *meta* positions through bridging benzene rings, thus labelled as *para*-NPG and *meta*-NPG, respectively. Our parameter-free, large-scale atomistic transport simulations show that within *para*-NPG, due to electronic coupling between GNRs, injected currents spread over a number of GNR channels, as reported for the recently fabricated NPG.[8,9] Contrary, within the *meta*-NPG, where QI takes place at bridging sites between GNRs, injected currents are confined within a single 0.7 nm wide GNR channel for up to distances larger than 100 nm. Moreover, this behavior is robust to out-of-plane distortions of benzene bridges,[27] which only have a minor effect for currents injected in the valence band. Importantly, recent bottom-up fabricated pairs of GNRs have been covalently ensembled with *para* and *meta* benzene-based connections (**Fig. 1d**),[15] which underscores the experimental feasibility of our proposed NPG structures. Finally, we demonstrate that the electronic tunability of *para* and *meta* connections allows designing a hybrid *meta-para-meta*-NPG where complex electrical paths are realized, with the additional benefit of being externally controllable via electrostatic gates.

QI is fundamentally correlated with π-conjugation[12] and so this work provides a simple, yet powerful, general toolkit for the design of bottom-up constructed carbon nanostructures where electron pathways may be quantum-engineered with nanometer, or sub-nanometer, precision. This, we believe, paves the way towards future carbon-based nanocircuitry.

## Methods

The DFT electronic structure and optimized geometries for the *para* and *meta*-NPG unit cells are obtained using SIESTA.[19,28] The unit cells are orthogonal and periodic, with cell parameters $a_x$ = 4.0 nm and $a_y$ = 0.8 nm, and contain 92 carbon atoms and 28 hydrogen atoms. We use a single-ζ basis set with 0.01 Ry energy shift. This choice neglects the existence of possible super-atom bands, which may be captured by more accurate basis sets.[8,29] However, it guarantees enough accuracy in the energy range ±2eV from $E_F$. We use norm-conserving Troullier-Martins pseudopotentials with a mesh cutoff of 400 Ry, and the GGA-PBE exchange-correlation functional.[30] The Brillouin zone is sampled using a 15 × 51 Monkhorst-Pack k-point mesh of the primitive NPG unit cell. Structural optimization is performed using a force threshold of 0.01 eV/Å.

The parameters for the TB models are obtained directly from the converged DFT Hamiltonian and overlap matrices[31] and correspond to all on-site and coupling elements associated with the carbon $p_z$ orbitals for the planar NPG structures, and s, $p_x$, $p_y$ and $p_z$ orbitals for the non-planar NPG structures. This model, obtained using the Python-based SISL utility,[18] retains the interaction range of the DFT basis set, is non-orthogonal, and takes the self-consistent effects of the hydrogen passivation into account. Transport calculations are then performed using this DFT-parameterized TB model and the Green's function (GF) formalism,[20,21,32] as implemented in TBtrans.[19] Transmission along transverse (longitudinal) direction $x$ ($y$) is obtained by repeating the NPG unit cell three times along each direction (see Supporting Fig. S5) and using 301 (71) k-points. For the large-scale TB & GF calculations we generate NPG supercells containing up to 345,920 atoms and covering areas up to 80 nm × 160 nm. We do not employ periodic boundary conditions in these large-scale calculations. Current injection is simulated via a constant, on-site level broadening $i\Gamma$ self-energy in the device Green's function,[33] localized on a single atom indicated with a red dot at $y$ = 0 in each device. This is equivalent as injecting electrons by means of an STM tip modelled with DFT/TB multi-scale simulations.[9] The value of $\Gamma$ acts mainly as a scaling factor for injected currents, and it is arbitrarily set to 1 eV. Currents are drained along the $y$ direction (longitudinal to GNR channels) into two semi-infinite NPG-like electrodes placed at the two terminations of the device along y, and absorbed by 5 nm wide regions at the left and right extremes equipped with complex absorbing potentials (CAP).[34,35] This avoids electronic back reflection due to the non-periodic walls of the device. Following Ref. 30, we visualize current flow by plotting 2D bond-current maps, summing all bond-current values flowing out of each atom (only positive-valued bond currents are considered). The colour-map is scaled in proportion to the current magnitude, so that areas with low to zero currents appear in dark purple. The colour range is always normalized to the maximum value of current.

## ASSOCIATED CONTENT

**Supporting Information**.
Fig. S1 – Out-of-plane conformations for *para* and *meta* NPGs
Fig. S2 – DFT and DFT-parametrized TB band structure comparison
Fig. S3 – Inter-channel coupling coefficient
Fig. S4 – DFT and DFT-parametrized TB band structure comparison for the out-of-plane conformations
Fig. S5 – Zero-bias transmission plots
Fig. S6 – Zero-bias transmission plots for the out-of-plane conformations
Fig. S7 – Bond currents at different energies
Fig. S8 – Bond currents at different energies for the out-of-plane conformations
Fig. S9 – DFT-parametrized TB and 1st nearest-neighbor band structure comparison
Fig. S10 – Atomic structures of interfaces used in the hybrid-NPGs
This material is available free of charge via the Internet at http://pubs.acs.org.

## AUTHOR INFORMATION

**Corresponding Author**
* E-mail I. A. R.: ialcon8@gmail.com

* E-mail M. B.: mabr@dtu.dk

**Present Addresses**

† Dipartimento di Ingegneria dell'Informazione, Università di Pisa, 56122 Pisa, Italy

§ Institut für Chemie und Biochemie - Physikalische und Theoretische Chemie, Freie Universität Berlin, Takustr. 3 14195 Berlin

**Author Contributions**

‡These authors contributed equally.

## ACKNOWLEDGMENT

We acknowledge useful conversations with Drs A. Garcia-Lekue and B. Kretz. Financial support by the Independent Research Fund Denmark (4184-00030) and Villum Fonden (00013340) is gratefully acknowledged. The Center for Nanostructured Graphene (CNG) is sponsored by the Danish National Research Foundation (DNRF103).

Insert Table of Contents artwork here

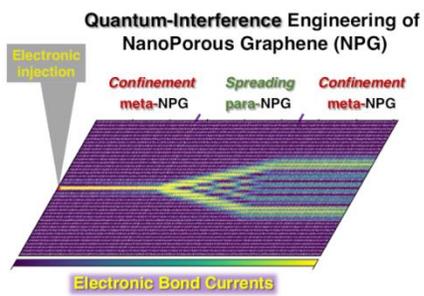